\documentclass[journal]{IEEEtran}
\IEEEoverridecommandlockouts

\usepackage{epsfig,rotating,setspace,latexsym,amsmath,epsf,amssymb,amsfonts,bm,theorem,cite,algorithm,graphicx,epsf,authblk,epstopdf,color,algpseudocode,bbm,subcaption}

\newtheorem{definition}{Definition}
\newtheorem{theorem}{Theorem}
\newtheorem{lemma}{Lemma}

\allowdisplaybreaks

\begin{document}

\title{Timely Status Updating Over Erasure Channels Using an Energy Harvesting Sensor:\\Single and Multiple Sources\thanks{This article has been presented in part at the 2018 Allerton Conference on Communication, Control, and Computing and at the 2019 IEEE International Symposium of Information Theory (ISIT).}\thanks{A. Arafa is with the Department of Electrical and Computer Engineering, University of North Carolina at Charlotte, Charlotte, NC 28223 USA (e-mail: aarafa@uncc.edu).}\thanks{J. Yang is with the School of Electrical Engineering and Computer Science, The Pennsylvania State University, State College, PA 16802 USA (e-mail: yangjing@psu.edu).}\thanks{S. Ulukus is with the Department of Electrical and Computer Engineering, University of Maryland at College Park, College Park, MD 20742 USA (e-mail: ulukus@umd.edu).}\thanks{H. V. Poor is with the Department of Electrical and Computer Engineering, Princeton University, Princeton, NJ 08544 USA (e-mail: poor@princeton.edu).}}

\author{Ahmed Arafa,~\IEEEmembership{Member,~IEEE}, Jing Yang,~\IEEEmembership{Member,~IEEE},\\ Sennur Ulukus,~\IEEEmembership{Fellow,~IEEE}, and H. Vincent Poor,~\IEEEmembership{Fellow,~IEEE}}

\maketitle

\begin{abstract}
A status updating system is considered in which data from multiple sources are sampled by an energy harvesting sensor and transmitted to a remote destination through an erasure channel. The goal is to deliver status updates of all sources in a timely manner, such that the cumulative long-term average {\it age-of-information} (AoI) is minimized. The AoI for each source is defined as the time elapsed since the generation time of the latest successful status update received at the destination from that source. Transmissions are subject to energy availability, which arrives in units according to a Poisson process, with each energy unit capable of carrying out one transmission from only one source. The sensor is equipped with a unit-sized battery to save the incoming energy. A scheduling policy is designed in order to determine which source is sampled using the available energy. The problem is studied in two main settings: no erasure status feedback, and perfect instantaneous feedback. 

For the case of one source, it is shown that {\it renewal policies} are optimal, in which successful status update instances form a renewal process. In the setting without feedback, it is further shown that {\it threshold-based policies} are optimal, in which the source is sampled only if the time until a new energy unit arrives exceeds a certain threshold. In the setting with feedback, threshold-greedy policies are investigated, in which the source is sampled according to a threshold-based policy following successful transmissions, and instantaneously whenever energy is available following failed transmissions. The optimal thresholds are found in closed-form in terms of the erasure probability. Such threshold-based policies are then extended for the case of multiple sources, combined with {\it round robin} scheduling in the setting without feedback, in which sources are sampled in the same repeating order; and {\it maximum-age-first} scheduling in the setting with feedback, in which sources with maximum AoI are given priority. In both settings, the achieved cumulative long-term average AoI is derived in closed-form in terms of the threshold, the erasure probability and the number of sources.
\end{abstract}

\section{Introduction}

Real-time sensing, monitoring and updating of physical phenomena is a main component of cyber-physical systems, 5G and beyond wireless systems, and internet-of-things (IoT) applications including industrial IoT (IIoT). Delivering {\it fresh} data is crucial in such systems so that optimal decisions are taken in a {\it timely} manner to maintain desirable system performance. The {\it age-of-information} (AoI) metric has been introduced in the literature for assessing the freshness and timeliness of data \cite{yates_age_1}, and is simple enough for implementation on low-complexity sensors, such as in IoT applications. When sensors rely on energy harvested from nature to communicate, it becomes essential to optimally manage the available energy to deliver timely data without risking running energy-hungry for long periods. In this work, we focus on timely status updating from multiple time-varying sources of data using a {\it shared} energy harvesting sensor over a noisy communication channel. We develop optimal transmission and scheduling policies that deliver status updates in a timely manner (with minimal AoI) subject to the availability of energy.

The AoI metric has been studied in the literature under various settings; mainly through modeling the update system as a queuing system and analyzing the long-term average AoI, and through using optimization tools to characterize optimal status updating policies, see, e.g., the recent survey in \cite{aoi-survey-jsac}.

In this paper, we consider a multiple source system monitored remotely through the help of data sent by an energy harvesting sensor. Energy arrives in units according to a Poisson process of unit rate, with each energy unit capable of only one transmission from only one source. Transmissions are composed of time-stamped packets (status updates) and are delivered to the remote destination through an erasure channel. Specifically, each status update is either erased with some probability or delivered instantly. With the goal of minimizing the cumulative long-term average AoI, we devise transmission and scheduling policies in two main settings regarding whether or not the sensor receives erasure feedback. 

We first focus on the case of one source, and show that {\it renewal policies} are optimal, in which successful status update times form a renewal process. We then show that optimal renewal policies admit a {\it threshold structure,} in which a new status update is transmitted only if the time until the next energy arrival since the latest transmission exceeds a certain threshold. In the system with feedback, this is complemented with {\it greedy} re-transmissions in case of failures, in which new status updates are transmitted whenever energy is available. The optimal thresholds are derived in closed-form in terms of the erasure probability. We then extend this to the case of multiple sources. We focus on threshold-based policies combined with {\it round robin} scheduling in the setting without erasure feedback, and {\it maximum-age-first} scheduling in the setting with feedback. Closed-form expressions for the AoI are derived in both cases. Several numerical results are presented to corroborate our theoretical findings.

\subsection{Related Works}

There have been a number of works focusing on analyzing AoI when transmitters rely on energy harvesting to communicate, e.g., \cite{yates_age_eh, elif_age_eh, liu-age-eh-sensing, arafa-age-2hop, arafa-aoi-eh-2hop-inf-battery, arafa-age-var-serv, elif-age-Emax, jing-age-online, jing-age-error-infinite-no-fb, jing-age-error-infinite-w-fb, jing-age-erasures-infinite-jour, baknina-age-coding, shahab-age-online-rndm, farazi-aoi-eh-preempt, zheng-aoi-eh-queue, baknina-updt-info, arafa-age-rbr, arafa-age-sgl, arafa-age-online-finite, elif-age-online-threshold, bacinoglu-aoi-eh-finite-gnrl-pnlty, arafa-age-erasure-no-fb, arafa-age-erasure-fb, krikidis-aoi-wpt, chen-aoi-eh-mac-het, leng-aoi-eh-cog-radio, stamatakis-aoi-eh-alarm, rafiee-aoi-eh-drop, ozel-aoi-eh-sensing, dong-aoi-mse, hirosawa2020aoiEH, jaiswal2020aoiEH, ko2020aoiEH, nouri2020aoiEH, abdelmagid2020aoiWPT, sleem2020aoiWPT, gindullina2020aoiEH, doncel2021age}. In summary, these works may be categorized according to the following three differentiating aspects: (1) the battery size, which can either be {\it finite} or {\it infinitely} large; (2) the energy arrival process knowledge, which can either be {\it offline}, i.e., predictable before the energy arrives (is harvested), or {\it online}, i.e., can only be known causally after the energy arrives; and (3) the service time, denoting the time for an update to traverse through the communication channel and reach the destination -- this can take multiple forms, but is mainly categorized into {\it deterministic} (zero or non-zero) services times, and {\it stochastic} service times. Our work in this paper can be categorized along the {\it finite battery, online, and deterministic service time} category. In addition, we consider the case of multiple sources over an erasure channel.

This paper includes extensions of the works in \cite{arafa-age-erasure-no-fb, arafa-age-erasure-fb} to the case of multiple sources. Other works in the literature focusing on channel erasures without energy harvesting constraints include, e.g., \cite{yates-age-erase-code, srivastava2019minimizing, javani2020age}.

Studying multiple sources with energy harvesting has been considered in \cite{hirosawa2020aoiEH, jaiswal2020aoiEH, gindullina2020aoiEH}. Reference \cite{hirosawa2020aoiEH} focuses on TDMA and FDMA schemes with average energy harvesting analysis, and provides means to choose between the two schemes given the available resources. Reference \cite{jaiswal2020aoiEH} is closely-related to the setting with feedback in our work. Following an MDP framework in a discrete-time setting with finite time horizon, the optimal policy is such that the sensor first probes the channel if the maximum AoI grows above a certain threshold, and then decides on sampling the source with maximum AoI if the probed channel conditions are better than a certain threshold as well. Different from \cite{jaiswal2020aoiEH}, we consider an infinite time horizon setting, differentiate between the setting with feedback and that without feedback, and provide analytical expressions for the AoI and the threshold under Poisson energy arrivals. Finally, the work in \cite{gindullina2020aoiEH} considers the notion of source diversity when multiple sources monitor the same physical phenomenon with different costs, and casts the optimal sampling problem as an MDP in a discrete-time setting.

\section{System Model}

We consider a system of $M$ sources of time-varying data that need to be monitored at a remote destination. At a given point in time, a sensor node chooses one of the sources, measures its status, time-stamps it, and sends a status update to the destination.\footnote{This setting is known in the literature as the {\it generate-at-will} model, owing to the fact that data from a chosen source can be generated when requested.} The sensor relies on energy harvested from nature to communicate with the destination. Energy arrives in units according to a Poisson process of unit (normalized) rate. Energy storage and energy expenditure are also normalized; the sensor is equipped with a unit-sized battery to save its harvested energy, and one status update transmission consumes one energy unit, i.e., completely depletes the battery. Any energy units arriving to a full battery are lost. In this setting, all sources share the same energy harvesting sensor, and hence {\it one energy unit can only update one source.}\footnote{In the study of energy harvesting communications, several works aim to take practical battery characteristics, e.g., leakage, inefficiency and lifetime, into the optimization framework, and investigate their impacts on the optimal energy management policies and system performances, see, e.g., \cite{gunduz_jcn, tcom/MichelusiBCCZ13, jsac/TutuncuogluYU15}. Such considerations are beyond the scope of our paper, and we defer it to future works on the subject.}

Status updates are subject to {\it erasures.} Specifically, the communication channel between the sensor and the destination is modeled as a time-invariant noisy channel, in which each update transmission gets erased with probability $q\in(0,1)$, independently from other transmissions. We differentiate between two main cases in our treatment:
\begin{enumerate}
\item {\it No updating feedback.} In this case, the sensor has no knowledge of whether an update is successful. Therefore, it can only use the up-to-date energy arrival profile and status updating decisions as well as the statistical information, such as the energy arrival rate and the erasure probability of the channel, to decide on which source to update next, and on the upcoming updating time points.
\item {\it Perfect updating feedback}. In this case, the sensor receives an instantaneous, error-free, feedback when an update is transmitted. Therefore, it decides what to do next based on the feedback information, along with the information it uses for the no feedback case.
\end{enumerate}

\begin{figure}[t]
\centering
\includegraphics[scale=.4]{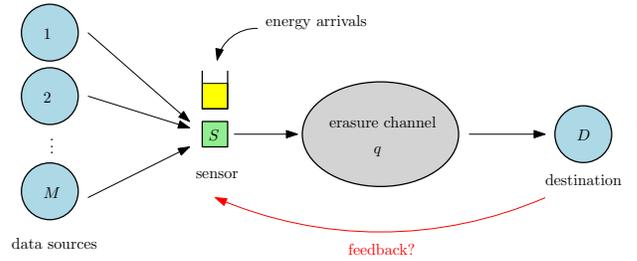}
\caption{System model.}
\label{fig:sys-mod}
\end{figure}

A complete system model describing the above features is shown in  Fig.~\ref{fig:sys-mod}.

Each status update carries a time stamp denoting when it was acquired at the sensor. From the destination's perspective, the AoI of source $j$ at time $t$, $a_j(t)$, is defined as the time elapsed since the latest update of source $j$ has been {\it successfully} received, i.e., with no erasures. This is mathematically given by
\begin{align}
a_j(t)=t-u_j(t),
\end{align}
where $u_j(t)$ is the time stamp of the last update of source $j$ that has been successfully received prior to time $t$.

Since each update transmission is not necessarily successful, we denote by $\{l_i\}$ the set of update transmission times, and by $\{s_i\}$ the times of the {\it successful} ones. Therefore, in general,  $\{s_i\}\subseteq\{l_i\}$. Let $\mathcal{E}(t)$ denote the amount of energy in the battery at time $t$. We then have the following energy causality constraint:
\begin{align} \label{eq:en-caus}
\mathcal{E}\left(l_i^-\right)\geq1,\quad\forall i,
\end{align}
where we use $l_i^-$ to denote the time instance right before $l_i$.

We assume that we begin with an empty battery at time $0$. The battery evolves as follows:
\begin{align} \label{eq:battery-inc}
\mathcal{E}\left(l_i^-\right)=\min\{\mathcal{E}\left(l_{i-1}^-\right)-1+\mathcal{A}(x_i),1\},\quad\forall i,
\end{align}
where $x_i\triangleq l_i-l_{i-1}$ denotes the inter-update attempt delay. We assume $s_0=l_0=0$ without loss of generality, i.e., the system starts with fresh information at time $0$. We denote by $\mathcal{F}_q$, the set of feasible transmission times $\{l_i\}$ described by (\ref{eq:en-caus}) and (\ref{eq:battery-inc}) in addition to an empty battery at time $0$, i.e., $\mathcal{E}(0)=0$.

\begin{figure}[t]
\center
\includegraphics[scale=.4]{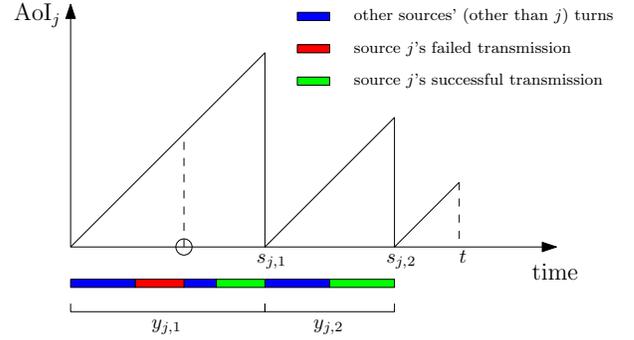}
\caption{Age evolution for source $j$ versus time with $n_j(t)=2$ successful updates. In this example, the first update is successfully received after two update attempts (a circle denotes failure).}
\label{fig:age_xmpl_erasure}
\end{figure}

Let us define $s_{j,i}$ as the time of the $i$th successful update pertaining to source $j$. Clearly, $\{s_{j,i}\}\subseteq\{s_i\},~\forall j\in[M]$. Further, let us denote by $y_{j,i}\triangleq s_{j,i}-s_{j,i-1}$ the {\it successful} inter-update delay of source $j$, and by $n_j(t)$ the number of updates from source $j$ that are {\it successfully} received by time $t$. We are interested in the average AoI given by the area under the age evolution curve of source $j$, see Fig.~\ref{fig:age_xmpl_erasure}, which is given by
\begin{align} \label{eq_aoi_area}
r_j(t)=\frac{1}{2}\sum_{i=1}^{n_j(t)}y_{j,i}^2+\frac{1}{2}\left(t-s_{j,n_j(t)}\right)^2.
\end{align}

The goal is to choose $(i)$ a set of feasible transmission times $\{l_1,l_2,l_3,\dots\}\in\mathcal{F}_q$, or equivalently $\{x_1,x_2,x_3,\dots\}$, and $(ii)$ a source scheduling policy $\pi$ to determine which source gets sampled at each transmission time, such that the {\it cumulative} long-term average AoI for all sources is minimized. That is, to solve the following optimization problem:
\begin{align} \label{opt_main}
\rho_{q,M}^\omega\triangleq\min_{\{x_i\}\in\mathcal{F}_q,\pi}~\limsup_{T\rightarrow\infty}\frac{1}{M}\sum_{j=1}^M\frac{1}{T}\mathbb{E}\left[r_j(T)\right],
\end{align}
where the expectation is taken over the joint distribution of all the underlying random variables, the superscript $\omega\equiv\text{noFB}$ in the case without updating feedback, and $\omega\equiv\text{wFB}$ in the case with perfect feedback.

We discuss the solution of problem (\ref{opt_main}) over the next two sections.

\section{The Single Source Case}

In this section, we solve problem (\ref{opt_main}) for $M=1$ source. In this case, we drop the subscript $j$ from the relevant expressions, and problem (\ref{opt_main}) reduces to only characterizing the optimal $\{x_i\}$.

\subsection{No Updating Feedback}

We first consider the case $\omega\equiv\text{noFB}$. We start by discussing a key characteristic of the optimal solution. Specifically, we show that the optimal status update policy is a renewal policy, in which the {\it actual} inter-update times $y_i$'s are independent and identically distributed (i.i.d.) and that the {\it actual} update times $s_i$'s form a renewal process. 

\subsubsection{Optimality of Renewal Policies}

\begin{figure}[t]
\center
\includegraphics[scale=0.75]{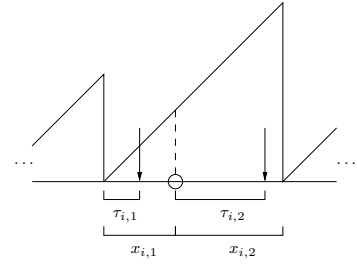}
\caption{AoI in the $i$th epoch with two update attempts. Arrows represent energy arrivals, and the circle denotes a failed update attempt.}
\label{fig:no_fb_age_epoch_tau}
\end{figure}

We first define some terminologies and notations. We use the term {\it epoch} to denote the time in between two consecutive successful updates. For instance, the $i$th epoch starts at time $s_{i-1}$ and ends at $s_i$, and has a length of $y_i$ time units. Note that an epoch may contain more than one update attempt, and the number of update attempts may vary from one epoch to another. We now slightly change our notation to fit it into our epoch definition and denote by $x_{i,k}$ the time in between the $(k-1)$th and the $k$th update attempt in the $i$th epoch. Similarly, let $\tau_{i,k}$ denote the time until the $k$th energy arrival in the $i$th epoch {\it starting from the $(k-1)$th update attempt}. For example, the first energy arrival in the $i$th epoch occurs at $s_{i-1}+\tau_{i,1}$, after which an update attempt occurs at $s_{i-1}+x_{i,1}$, with $x_{i,1}\geq\tau_{i,1}$ due to energy causality (\ref{eq:en-caus}). Now say that this first update attempt has failed. Then, the sensor waits for the second energy arrival in the epoch occurring at $s_{i-1}+x_{i,1}+\tau_{i,2}$, after which the second update attempt occurs at $s_{i-1}+x_{i,1}+x_{i,2}$, and so on. Note that according to the definition of $\tau_{i,k}$'s, they do not necessarily represent the energy inter-arrival times, since $x_{i,k}$ can be strictly larger than $\tau_{i,k}$ (see Fig.~\ref{fig:no_fb_age_epoch_tau}).

Observe that transmission attempts occurring in the $i$th epoch may depend, in principal, on the history of events (transmission attempts and energy arrivals) that had occurred before the epoch started, which we denote by $\mathcal{H}_{i-1}$. Theorem~\ref{thm:rnwl_noFB} below shows, under some regularity conditions, that this is not the case; events in an epoch are independent of the history of events in previous epochs. Before we make that statement precise, we focus on the following special case of online policies, which are also the focus in \cite{jing-age-online, arafa-age-online-finite}:
\begin{definition}[Uniformly Bounded Policy] \label{def_ubp}
An online policy whose inter-update times have a bounded second moment is called uniformly bounded.
\end{definition}
Intuitively, one would expect practical status update policies to be uniformly bounded as per Definition~\ref{def_ubp}, so that the inter-update delays do not grow infinitely large (in expectation). We now state the main result of this section in Theorem~\ref{thm:rnwl_noFB} below. The proof of the theorem, which is fully presented in Appendix~\ref{apndx_thm:rnwl_noFB}, is similar in essence to the proofs of \cite[Theorems 1 and 2]{arafa-age-online-finite} albeit some notable differences.

\begin{theorem} \label{thm:rnwl_noFB}
In the optimal solution of problem (\ref{opt_main}) with $M=1$ and $\omega\equiv\text{noFB}$, any uniformly bounded policy is outperformed by a renewal policy in which the epoch lengths, $y_i$'s, are i.i.d.
\end{theorem}

\subsubsection{Optimal Renewal Policy: Threshold Structure}

We now analyze the best renewal policy and show that it has a {\it threshold} structure. Theorem~\ref{thm:rnwl_noFB} shows that epoch starting times, $s_i$'s, at which the system resets by making both the sensor's battery and the AoI drop to $0$ simultaneously, constitute a renewal process. Since epoch lengths are i.i.d., we drop the subscript $i$ from all random variables and denote the epoch duration by $y$ and the inter-update attempt duration by $x$. {\it Observe that we do not differentiate between different update attempts in a single epoch since the sensor is unaware of this information due to lack of erasure feedback.} From the sensor's point of view, it only designs a single inter-update attempt duration $x$. However, it takes the value of $q$ into account while doing so as we show in the sequel. Let us (re)define $\tau$ as the time elapsed until the next energy arrival starting from the previous update attempt. Given that the sensor is unaware of erasure events, the sensor is ignorant of when an epoch starts or ends; the only available information to base its next update instant is the time $\tau$. In other words, inter-update attempt times in the epoch are functions of only the most recent energy arrival time; that is, $x$ is only a function of $\tau$.

By the strong law of large numbers for renewal processes (the renewal reward theorem) \cite{ross_stochastic}, problem (\ref{opt_main}) with $M=1$ and $\omega\equiv\text{noFB}$ now reduces to an optimization over a single epoch as follows:
\begin{align} \label{opt_no_fb_rnwl}
\rho_{q,1}^{\text{noFB}}=\min_{x(\cdot)}\quad&\frac{\mathbb{E}\left[R\right]}{\mathbb{E}\left[y\right]} \nonumber \\
\mbox{s.t.}\quad&x(\tau)\geq\tau,\quad\forall\tau,
\end{align}
where $R$ denotes the area under the AoI curve (the reward) in the epoch. We now introduce the following auxiliary problem to solve the one above:
\begin{align} \label{opt_no_fb_aux}
p^{\text{noFB}}(\lambda)\triangleq\min_{x(\cdot)}\quad&\mathbb{E}\left[R\right]-\lambda\mathbb{E}\left[y\right] \nonumber \\
\mbox{s.t.}\quad&x(\tau)\geq\tau,\quad\forall\tau,
\end{align}
for some $\lambda\geq0$. One can show that $\rho_{q,1}^{\text{noFB}}$ is given by $\lambda^*$ that solves $p^{\text{noFB}}(\lambda^*)=0$, and that such $\lambda^*$ is unique since $p^{\text{noFB}}(\lambda)$ is decreasing in $\lambda$ \cite{dinkelbach-fractional-prog}. The next theorem characterizes the solution of problem (\ref{opt_no_fb_aux}). The proof is in Appendix~\ref{apndx_thm:threshold_noFB}.

\begin{theorem} \label{thm:threshold_noFB}
The optimal solution of problem (\ref{opt_no_fb_aux}) depends on $q$. If $q<\frac{1}{2}$, then it is a $\lambda^\prime$-threshold policy, in which
\begin{align} \label{eq_no_fb_x_1}
x(t)=\begin{cases}\lambda^\prime,\quad &t<\lambda^\prime\\ t,\quad &t\geq\lambda^\prime \end{cases},
\end{align}
where $\lambda^\prime$ is the unique solution of 
\begin{align} \label{eq_no_fb_lmda_prm}
\frac{1+q}{1-q}\lambda^\prime+\frac{2q}{1-q}e^{-\lambda^\prime}=\lambda.
\end{align}
Otherwise, if $q\geq\frac{1}{2}$, then the optimal solution is greedy, in the sense that $x(t)=t~\forall t$.
\end{theorem}

We conclude this section by stating a few remarks. First, observe that for the case of no erasures, i.e., $q=0$, (originally considered in \cite{jing-age-online}) we get from (\ref{eq_no_fb_lmda_prm}) and (\ref{eq_no_fb_p}) that $\lambda^\prime=\lambda$ and $p(\lambda)=e^{-\lambda}-\frac{1}{2}\lambda^2$, respectively, coinciding with the optimal solution in \cite{jing-age-online}. Second, for a given $\lambda\geq0$, (\ref{eq_no_fb_lmda_prm}) shows that $\lambda^\prime\leq\lambda$ with equality if and only if $q=0$. This shows that the problem with erasures does {\it not} have the recurring property shown in \cite{jing-age-online, arafa-age-online-finite, elif-age-online-threshold} that the optimal long-term average AoI equals the optimal threshold; they are only equal if $q=0$.

\subsection{Perfect Updating Feedback}

We now consider the case $\omega\equiv\text{wFB}$. As in the no feedback case, we also begin by showing that renewal policies are optimal.

\subsubsection{Optimality of Renewal Policies}

We focus on the class of uniformly bounded policies (as per Definition~\ref{def_ubp}). The next theorem shows that renewal policies are optimal in that regard.

\begin{theorem} \label{thm:rnwl_fb}
In the optimal solution of problem (\ref{opt_main}) with $M=1$ and $\omega\equiv\text{wFB}$, any uniformly bounded policy is outperformed by a renewal policy in which the epoch lengths, $y_i$'s, are i.i.d.
\end{theorem}

The proof of the theorem goes along the same lines as in that of Theorem~\ref{thm:rnwl_noFB}. Specifically, we prove Theorem~\ref{thm:rnwl_noFB} by considering a genie-aided setup in which a genie informs the sensor of when its updates were successful, and then argue that in the optimal policy such genie's information can be discarded. One can slightly manipulate such arguments to prove Theorem~\ref{thm:rnwl_fb} above by treating the genie-aided system as exactly the feedback system considered in this section. The details of the proof are omitted for brevity.

\subsubsection{Threshold Greedy Policies}

Now that the optimality of renewal-type policies is established by Theorem~\ref{thm:rnwl_fb}, we proceed with characterizing the optimal renewal policy in this section. 

Since epoch lengths are i.i.d., by the strong law of large numbers for renewal processes (the renewal-reward theorem) \cite{ross_stochastic}, the objective function of problem (\ref{opt_main}) with $M=1$ and $\omega\equiv\text{wFB}$ is given by
\begin{align} \label{eq_obj_rnwl_fb}
\limsup_{T\rightarrow\infty}\frac{1}{T}\mathbb{E}\left[r(T)\right]=\frac{\mathbb{E}\left[R\left({\bm x}\right)\right]}{\mathbb{E}\left[y\left({\bm x}\right)\right]},
\end{align}
where $R$ denotes the area under the AoI curve (the reward) in the epoch, and ${\bm x}=\{x_1,x_2,\dots\}$ is the update policy within the epoch where $x_i$ now denotes the time elapsed from the {\it beginning of the epoch} until the $i$th update attempt\footnote{We slightly deviate from the original definition of $x_i$, and assume without loss of generality that the epoch starts at time $0$.}, and the expectation is taken with respect to the energy arrivals' distribution within the epoch. Different from the case with no updating feedback, we emphasize the dependency of $R$ and $y$ on the updating policy ${\bm x}$, since now the sensor receives a feedback for each update attempt. We show later on the effect of this feedback and how it can be best employed. 

\begin{figure}
\center
\includegraphics[scale=.75]{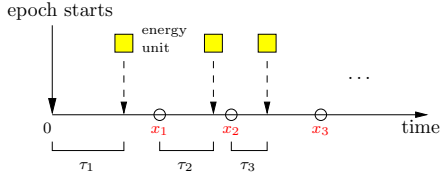}
\caption{Illustration of the notations used to describe energy arrivals and update attempt times within the epoch, in the setting with perfect updating feedback.}
\label{fig:tau_epoch_fb}
\end{figure}

Let $\tau_1$ denote the time until the first energy arrival in the epoch, and $\tau_i$, $i\geq2$, denote the time until energy arrives {\it after} the $i$th update attempt, i.e., after time $x_i$, see Fig.~\ref{fig:tau_epoch_fb}. We now have the following lemma:

\begin{lemma} \label{thm:policy_depends_aoi}
In the optimal policy, $x_i$ only depends on the AoI at $\tau_i+x_{i-1}$, i.e., $x_i\equiv x_i\left(a\left(\tau_i+x_{i-1}\right)\right)$, with $x_0\triangleq0$.
\end{lemma}
The proof of Lemma~\ref{thm:policy_depends_aoi} mainly depends on the memoryless property of the exponential distribution, along the same lines of the proof of \cite[Lemma~3]{arafa-age-online-finite}, and is omitted for brevity. By Lemma~\ref{thm:policy_depends_aoi}, we have $x_1\equiv x_1\left(\tau_1\right)$, $x_2\equiv x_2\left(\tau_2+x_1(\tau_1)\right)$, $x_3\equiv x_3\left(\tau_3+x_2\left(\tau_2+x_1(\tau_1)\right)\right)$, and so on.

By (\ref{eq_obj_rnwl_fb}) and Lemma~\ref{thm:policy_depends_aoi}, problem (\ref{opt_main}) with $M=1$ and $\omega\equiv\text{wFB}$ now reduces to an optimization over a single epoch as follows:
\begin{align} \label{opt_epoch_fb}
\min_{{\bm x}}\quad&\frac{\mathbb{E}\left[R\left({\bm x}\right)\right]}{\mathbb{E}\left[y\left({\bm x}\right)\right]} \nonumber \\
\mbox{s.t.}\quad&x_1\left(\tau_1\right)\geq\tau_1 \nonumber \\
&x_2\left(\tau_2+x_1\left(\tau_1\right)\right)\geq\tau_2+x_1\left(\tau_1\right) \nonumber \\
&x_3\left(\tau_3+x_2\left(\tau_2+x_1\left(\tau_1\right)\right)\right)\geq\tau_3+x_2\left(\tau_2+x_1\left(\tau_1\right)\right) \nonumber \\
&\dots,
\end{align}
where the inequalities represent energy causality constraints. Using iterated expectations on the (independent) erasure events, $\mathbb{E}\left[R\left({\bm x}\right)\right]$ is given by
\begin{align}
\mathbb{E}\left[R\left({\bm x}\right)\right]=&(1-q)\frac{1}{2}\mathbb{E}\left[x_1^2\left(\tau_1\right)\right] \nonumber \\
&+q(1-q)\frac{1}{2}\mathbb{E}\left[x_2^2\left(\tau_2+x_1\left(\tau_1\right)\right)\right] \nonumber \\
&+q^2(1-q)\frac{1}{2}\mathbb{E}\left[x_3^2\left(\tau_3+x_2\left(\tau_2+x_1\left(\tau_1\right)\right)\right)\right] \nonumber \\
&+\dots,
\end{align}
with $\mathbb{E}\left[y\left({\bm x}\right)\right]$ given similarly as above after excluding the $\frac{1}{2}$ terms and the squaring of the $x_i$'s.

As in the case with no updating feedback, we introduce the following auxilliary problem to get a handle on problem (\ref{opt_epoch_fb}):
\begin{align} \label{opt_aux_fb}
p^{\text{wFB}}\left(\lambda\right)\triangleq\min_{{\bm x}}\quad&\mathbb{E}\left[R\left({\bm x}\right)\right]-\lambda\mathbb{E}\left[y\left({\bm x}\right)\right] \nonumber \\
\mbox{s.t.}\quad&\text{problem (\ref{opt_epoch_fb})'s constraints},
\end{align}
with $\lambda\geq0$. As before, one can show that $\rho_{q,1}^{\text{wFB}}$ is given by $\lambda^*$ that solves $p^{\text{wFB}}(\lambda^*)=0$, and that such $\lambda^*$ is unique since $p^{\text{wFB}}(\lambda)$ is decreasing in $\lambda$ \cite{dinkelbach-fractional-prog}. 

We now focus on characterizing $p^{\text{wFB}}(\lambda)$. Towards that end, we use two terminologies in order to refer to the structure of $x_i$, for any $i$. We call $x_i$ a {\it greedy policy} if the $i$th update attempt in the epoch takes place immediately after $\tau_i$. In this case, the constraint on $x_i$ (the $i$th lower bound constraint in problem (\ref{opt_epoch_fb})) is satisfied with equality. On the other hand, we call $x_i$ a {\it $\gamma$-threshold} policy if the $i$th update attempt in the epoch only takes effect if the AoI grows above $\gamma$, and $x_i(t)$ would be defined as in (\ref{eq_no_fb_x_1}) after replacing $\lambda^\prime$ with $\gamma$.
We now have the following lemma (we use the notation $[\cdot]^+\triangleq\max(\cdot,0)$; the proof of the lemma is in Appendix~\ref{apndx_thm:thrshld_grd_fb}):
\begin{lemma} \label{thm:thrshld_grd_fb}
In problem (\ref{opt_aux_fb}), the following two claims are equivalent:
\begin{itemize}
    \item[(A)] $x_1$ is a $\gamma$-threshold policy.
    \item[(B)] $\{x_i,~i\geq2\}$, are all greedy policies.
\end{itemize}
Furthermore, if Claim $(B)$ holds, then $\gamma=\left[\lambda-\frac{q}{1-q}\right]^+$.
\end{lemma}

We term the policies of Lemma~\ref{thm:thrshld_grd_fb} {\it threshold-greedy} policies. Employing such policies is quite intuitive in systems with feedback. Firstly, after an update is successfully transmitted, the AoI drops down to $0$. One should therefore wait for some time at least (the threshold $\gamma$ in this case) before attempting a new transmission. Such approach has been shown to be optimal in, e.g., \cite{jing-age-online, elif-age-online-threshold, arafa-age-online-finite}, in addition to the system without feedback in \cite{arafa-age-erasure-no-fb}. Secondly, if this new transmission attempt fails, then the AoI continues to increase until another energy unit arrives. It is therefore intuitive to update right away, i.e., greedily, after such energy unit arrives since the AoI is already high enough (higher than the threshold $\gamma$), and repeat that until the update is eventually successful. The above lemma shows that threshold-greedy policies are not just intuitive, but are actually representing a fixed-point solution of the problem. The next theorem characterizes the optimal threshold-greedy policy. The proof is in Appendix~\ref{apndx_thm:threshold_wFB}.

\begin{theorem} \label{thm:threshold_wFB}
The optimal threshold-greedy policy that solves problem (\ref{opt_aux_fb}) is such that $\gamma^*=\lambda^*-\frac{q}{1-q}>0$, with $\lambda^*$ being the unique solution of
\begin{align} \label{eq_lmda_fb}
e^{-\left(\lambda^*-\frac{q}{1-q}\right)}+\frac{2q-q^2}{2(1-q)^2}=\frac{1}{2}\left(\lambda^*\right)^2.
\end{align}
\end{theorem}

To summarize, given the erasure probability $q$, the optimal first status update policy (following a successful transmission) is a $\left(\lambda^*-\frac{q}{1-q}\right)$-threshold policy, and then all update attempts after the first one (following unsuccessful transmissions) are greedy. $\lambda^*$ is the unique solution of (\ref{eq_lmda_fb}), which also represents the long-term average AoI (the value of (\ref{eq_obj_rnwl_fb})).

\section{The Multiple Sources Case}

We now extend our solutions in the previous section to the case of $M\geq2$ sources. Inspired by their optimality in the single source case, we focus on renewal-type policies for the multiple sources case as well. Renewals here, however, are defined with respect to each source $j$. Specifically, we focus on updating policies in which the times in between two consecutive updates for {\it source $j$} are i.i.d. Such times are governed by the scheduling policy employed, which we describe in detail over the following subsections.

\subsection{No Updating Feedback}

In the case of no updating feedback, we focus on a {\it round robin} (RR) scheduling policy $\pi_\text{RR}$, in which the sensor samples sources in the order $1,2,\dots,M$ and repeats. Each incoming energy unit is assigned to sample and transmit the source whose turn comes up in such order. 

Given their optimality in the single source case, we focus on threshold-based policies, in which the sensor samples a source only if {\it the time until its assigned energy unit arrives surpasses a certain threshold $\gamma$.} Such threshold $\gamma$ is the same for all sources given the symmetric system conditions.

In what follows, we analyze the long-term average AoI of some source $j$ under $\pi_\text{RR}$ and $\gamma$-threshold policies. Clearly, under such policies, the long-term average AoI's of all sources become {\it identical,} which we denote $\tilde{\rho}_{q,M}^{\text{noFB}}\left(\text{RR},\gamma\right)$. We now have the following theorem (the proof is in Appendix~\ref{apndx_thm:multisrc_noFB}):
\begin{theorem}\label{thm:multisrc_noFB}
Consider problem (\ref{opt_main}) with $\omega\equiv\text{noFB}$. RR scheduling and $\gamma$-threshold policies achieve the following cumulative long-term average AoI:
\begin{align} \label{eq_rho_M_noFB}
\tilde{\rho}_{q,M}^{\text{noFB}}\left(\text{RR},\gamma\right)=&\frac{\frac{1}{2}\gamma^2+(\gamma+1)e^{-\gamma}}{\gamma+e^{-\gamma}} \nonumber \\ &+\left(\frac{M-1}{2}+\frac{Mq}{1-q}\right)\left(\gamma+e^{-\gamma}\right).
\end{align}
\end{theorem}

Based on Theorem~\ref{thm:multisrc_noFB}, one can find the optimal threshold $\gamma^*$ that minimizes $\tilde{\rho}_{q,M}^{\text{noFB}}\left(\text{RR},\gamma\right)$.

\subsection{Perfect Updating Feedback}

In the case of perfect updating feedback, we focus on a {\it maximum-age-first} (MAF) scheduling policy $\pi_{\text{MAF}}$, which we define next.

\begin{definition}[Maximum-Age-First (MAF) Scheduling]
The maximum-age-first (MAF) scheduling policy $\pi_{\text{MAF}}$ schedules the source with maximum AoI to be sampled when the next energy unit becomes available.
\end{definition}

MAF scheduling policies are intuitive since one focuses on minimizing the cumulative AoI of all sources. In addition, they have been shown optimal when channel conditions are symmetric across sources \cite{kadota-aoi-broadcast}, and also in \cite{bedewy-aoi-multisource} through a stochastic ordering argument when all sources incur the same age-penalty. Observe that our system model is symmetric since updates from all sources encounter the same channel with i.i.d. erasure events, and all sources incur the same normalized linear age-penalty (vanilla AoI). We note that MAF and RR are essentially the same if $q=0$. In case $q>0$ the two policies may differ. Also note that MAF scheduling requires knowledge of the AoI of each source at the destination, i.e., requires erasure feedback.

We combine $\pi_{\text{MAF}}$ with $\gamma$-threshold-greedy policies. That is, if source $j$ is to be sampled next according to $\pi_{\text{MAF}}$, it only gets sampled if the time until its assigned energy unit arrives surpasses a certain threshold $\gamma$, and {\it then in case of failure it follows that with greedy re-transmissions for the same source $j$ until successful reception.}

Next, we analyze the long-term average AoI of some source $j$ under $\pi_\text{MAF}$ and $\gamma$-threshold-greedy policies. As in the no updating feedback case, under such policies, the long-term average AoI's of all sources also become {\it identical,} which we denote $\tilde{\rho}_{q,M}^{\text{wFB}}\left(\text{MAF},\gamma\right)$. We now have the following theorem (the proof is in Appendix~\ref{apndx_thm:multisrc_wFB}):
\begin{theorem}\label{thm:multisrc_wFB}
Consider problem (\ref{opt_main}) with $\omega\equiv\text{wFB}$. MAF scheduling and $\gamma$-threshold-greedy policies achieve the following cumulative long-term average AoI:
\begin{align} \label{eq_rho_M_wFB}
\tilde{\rho}_{q,M}^{\text{wFB}}\left(\text{MAF},\gamma\right)=&\frac{\frac{1}{2}\gamma^2+(\gamma+1)e^{-\gamma}\left(\gamma+e^{-\gamma}\right)\frac{q}{1-q}+\frac{q}{(1-q)^2}}{\gamma+e^{-\gamma}+\frac{q}{1-q}}
\nonumber \\ 
&+\frac{M-1}{2}\left(\gamma+e^{-\gamma}+\frac{q}{1-q}\right).
\end{align}
\end{theorem}

Based on Theorem~\ref{thm:multisrc_wFB}, one can find the optimal threshold $\gamma^*$ that minimizes $\tilde{\rho}_{q,M}^{\text{wFB}}\left(\text{MAF},\gamma\right)$. We note that $\tilde{\rho}_{0,M}^{\text{noFB}}\left(\text{RR},\gamma\right)=\tilde{\rho}_{0,M}^{\text{wFB}}\left(\text{MAF},\gamma\right)$, i.e., the two cases (with and without feedback) admit the same cumulative long-term AoI expressions in the case of no erasures, as expected.

\section{Numerical Results}

In this section, we corroborate our analysis by presenting some numerical examples. 

\begin{figure}[t]
\center
\includegraphics[scale=.4]{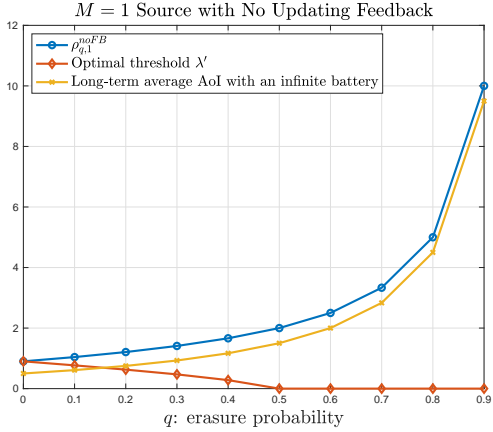}
\caption{Optimal AoI for $M=1$ source without feedback $\rho_{q,1}^{\text{noFB}}$, and that with an infinite battery \cite{jing-age-error-infinite-no-fb}, along with the optimal threshold $\lambda^\prime$, versus the erasure probability $q$.}
\label{fig_no_fb_age_q}
\end{figure}

We first start with the single source case. For the setting without feedback, we plot the optimal AoI derived in Theorem~\ref{thm:threshold_noFB} versus the erasure probability, along with the corresponding optimal threshold in Fig.~\ref{fig_no_fb_age_q}. For the case $q\leq\frac{1}{2}$, we basically start with a large-enough value of $\lambda^\prime$ that makes $p\left(\lambda^\prime\right)<0$, and then use a bisection search (in between $0$ and that large-enough value) to find $\lambda^\prime$ that solves $p\left(\lambda^\prime\right)=0$. We then use (\ref{eq_no_fb_lmda_prm}) to find the optimal long-term average AoI $\lambda^*=\rho_{q,1}^{\text{noFB}}$. We also plot the optimal long-term average AoI for the infinite battery case for comparison, which has been shown in \cite{jing-age-error-infinite-no-fb} to be equal to $\frac{1+q}{2(1-q)}$. Clearly, the solution for the infinite battery case serves as a lower bound for the solution for the $B=1$ case. From the figure, we also see that, quite intuitively, the larger the erasure probability, the larger the AoI, i.e., $\rho_{q,1}^{\text{noFB}}$ is monotonically increasing in $q$. In addition, we see that the optimal threshold $\lambda^\prime$ is monotonically decreasing in $q$. This is quite intuitive, since the sensor should be more eager to send new updates if the erasure probability is high, so that when the update is eventually received successfully the AoI would not be large.

\begin{figure}[t]
\center
\includegraphics[scale=.4]{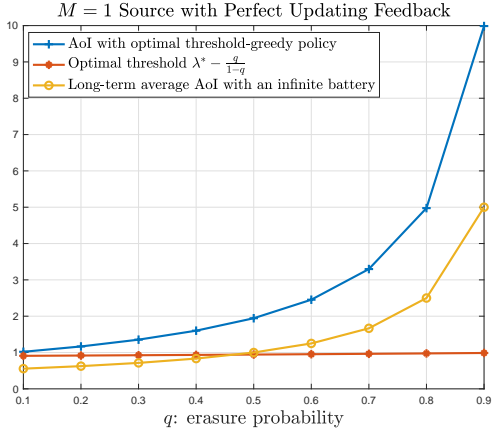}
\caption{AoI with optimal threshold-greedy policy for $M=1$ source with feedback, and that with an infinite battery \cite{jing-age-error-infinite-no-fb}, along with the optimal threshold $\lambda^*-\frac{q}{1-q}$, vs.~the erasure probability $q$.}
\label{fig_aoi_q}
\end{figure}

For the case with feedback, we plot the long-term average AoI achieved with the optimal threshold-greedy policy of Theorem~\ref{thm:threshold_wFB} versus the erasure probability $q$ in Fig.~\ref{fig_aoi_q}. We also plot the optimal threshold $\lambda^*-\frac{q}{1-q}$, and compare the results with that of the infinite battery case, derived in \cite{jing-age-erasures-infinite-jour} to be $\frac{1}{2(1-q)}$. We see that the AoI increases with $q$, which is quite expected. We also note that the optimal threshold is almost constant. This is attributed to the fact that as $q$ increases, both $q/(1-q)$ and $\lambda^*$ in (\ref{eq_lmda_fb}) increase by almost the same amount. 

\begin{figure}[t]
\center
\includegraphics[scale=.4]{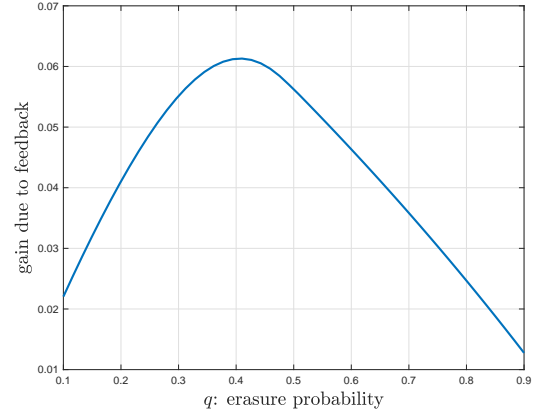}
\caption{Difference between the long-term average AoI without feedback and that with feedback versus the erasure probability $q$.}
\label{fig_feedback_gain}
\end{figure}

In Fig.~\ref{fig_feedback_gain}, we analyze the benefits of having a feedback link by plotting the difference between the long-term average AoI in the case without feedback and that with feedback versus the erasure probability $q$. We denote such difference by the {\it gain due to feedback} in the figure. We observe that the gain is highest around mid values of $q$, and decreases around its extremal values. The main reason behind this is that for relatively low values of $q$, the two systems (with and without feedback) are almost identical since erasures are not very common. While for relatively high values of $q$, feedback is not really helpful since erasures would occur more frequently anyway. It is in that mid range around $q=0.4$ that feedback makes a difference.

\begin{figure}[t]
\center
\includegraphics[scale=.4]{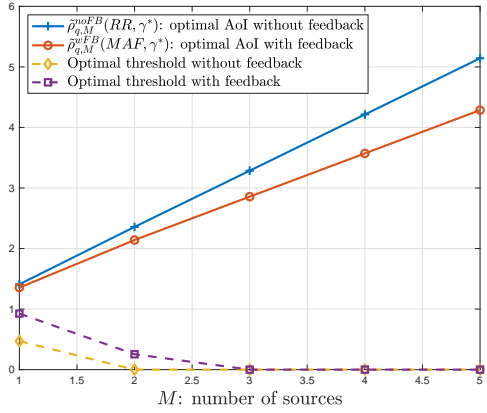}
\caption{Comparison of optimal AoI without feedback $\tilde{\rho}_{q,M}^{\text{noFB}}(\text{RR},\gamma^*)$ and that with feedback $\tilde{\rho}_{q,M}^{\text{wFB}}(\text{MAF},\gamma^*)$ versus the number of sources $M$, with erasure probability $q=0.3$.}
\label{fig_multi-src_q3}
\end{figure}

We now focus on the multiple sources case. Using the results of Theorem~\ref{thm:threshold_noFB} and Theorem~\ref{thm:threshold_wFB}, we numerically compute the optimal threshold $\gamma^*$ for different values of $M$, with fixed erasure probability $q=0.3$, and plot the results in Fig.~\ref{fig_multi-src_q3}. One can see that, as expected, the long-term average AoI for both settings of feedback is increasing with $M$, while the optimal threshold is decreasing. In addition, we see that greedy becomes optimal in case the number of sources exceeds $2$ (in the case without feedback) and $3$ (in the case with feedback). This relatively small number of critical sources after which greedy becomes optimal is mainly attributed to the usage of a unit battery at the sensor.

\begin{figure}[t]
\center
\includegraphics[scale=.4]{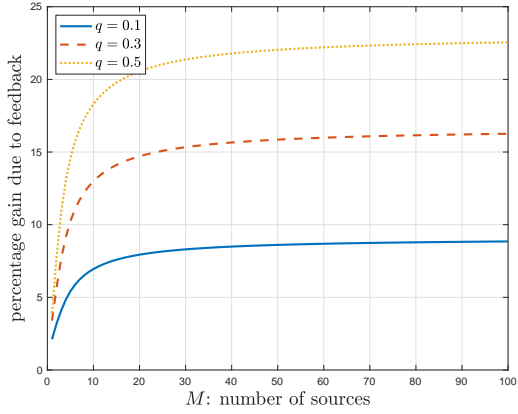}
\caption{Percentage gain due to feedback defined in (\ref{eq_percent_gain}) versus the number of sources $M$, with different values of the erasure probability $q$.}
\label{fig_percent-gain}
\end{figure}

Finally, in Fig.~\ref{fig_percent-gain}, we examine the behavior of our systems as the number of sources grows large. Specifically, we compute the {\it percentage gain due to feedback}, which we define as
\begin{align}\label{eq_percent_gain}
\left(1-\frac{\tilde{\rho}_{q,M}^{\text{wFB}}(\text{MAF},\gamma^*)}{\tilde{\rho}_{q,M}^{\text{noFB}}(\text{RR},\gamma^*)}\right)\times100\%,
\end{align}
and plot it against $M$ for different values of $q$. We see that the percentage gain converges to a specific $q$-dependent value as $M$ grows. This shows that while feedback does enhance the system's performance, the main performance is only dependent on the erasure probability $q$ for a large number of sources.

\section{Conclusions}

A multi-source status updating system has been considered, in which a shared energy harvesting sensor samples one source at a given time, subject to energy availability, and transmits a relevant status update to a remote destination over an erasure channel. Based on whether the feedback erasure status is available at the sensor, multiple transmission and scheduling policies have been derived to minimize the cumulative long-term average AoI. Our analysis has focused on erasure-dependent threshold-based policies, in which a new source is sampled only if the AoI grows above a certain threshold that depends on the erasure probability. Expressions for the optimal thresholds and the corresponding AoI's have been derived, and numerous corroborating numerical results have been presented.

Future work includes extending the results of this paper to the case of arbitrary-sized batteries, with more involved battery storage and leakage models, as well as providing analytical proofs of optimality for RR (MAF) scheduling for the case without (with) erasure status feedback.

\appendix

\subsection{Proof of Theorem~\ref{thm:rnwl_noFB}} \label{apndx_thm:rnwl_noFB}

We follow an indirect approach to prove the theorem. Basically, we derive an achievable lower bound on the long-term average AoI using renewal-type policies in a {\it genie-aided} system in which there exists a genie that informs the sensor when updates are successful, i.e., the epochs' start times. However, we enforce a constraint on the sensor {\it not to use} the lack of this piece of information to infer that its update is unsuccessful and act accordingly to change its policy within the same epoch. This seemingly unintuitive constraint simplifies the proof as we will see later on. Now observe that such genie-aided system cannot perform worse than the original system that we consider in this paper, and hence, a lower bound on this genie-aided system is also a lower bound on the original one. We then conclude the proof by showing that such lower bound is also achievable in the original system by showing that the optimal renewal-type policy does not actually need the information provided by the genie, thereby proving optimality of renewal-type policies in the original system as well. Next, we provide the details.

In the genie-aided system, consider any online feasible uniformly bounded policy $\{x_{i,k}\}$. Focusing on the $i$th epoch, let us denote by $R_{i,m}$ the area under the age curve in the $i$th epoch given that it went through $m$ update attempts, and by $R_i$ the area under the age curve in it {\it irrespective} of how many update attempts. Let us also denote by $e_{i,k}$ the event that the $k$th update attempt in the $i$th epoch gets erased. We can now write the following:
\begin{align}
R_{i,m}=&\frac{1}{2}\left(x_{i,1}+x_{i,2}+\dots+x_{i,m}\right)^2, \\
R_i=&\sum_{m=1}^\infty R_{i,m}\cdot\prod_{k=1}^{m-1}\mathbbm{1}\left(e_{i,k}\right)\mathbbm{1}\left(e_{i,m}^c\right), \label{eq_r_rm}
\end{align}
where $\mathbbm{1}(\cdot)$ is the indicator function, and the superscript $c$ denotes the complement of an event.

Next, for a fixed time $T$, denote by $N_T$ the number of epochs that have already {\it started} by time $T$. Given the history before the $i$th epoch, $\mathcal{H}_{i-1}$, and the number of update attempts in the $i$th epoch, $m$, let us define the vector ${\bm \tau}_i^{(m)}\triangleq[\tau_{i,1},\tau_{i,2},\dots,\tau_{i,m}]$, and define the following statistical average of the area under the age curve in the $i$th epoch with $m$ update attempts:
\begin{align}
\hat{R}_{i,m}\left({\bm \gamma}^{(m)},\mathcal{H}_{i-1}\right)\triangleq\mathbb{E}\left[R_{i,m}\Big|{\bm \tau}_i^{(m)}={\bm \gamma}^{(m)},\mathcal{H}_{i-1}\right].
\end{align}
Therefore, it holds that
\begin{align} \label{eq_rm_itr_exp}
&\hspace{-.075in}\mathbb{E}\left[R_{i,m}\mathbbm{1}\left(i\leq N_T\right)\right] \nonumber \\
&\hspace{-.075in}=\mathbb{E}_{\mathcal{H}_{i-1}}\!\!\left[\mathbb{E}_{{\bm \tau}_i^{(m)}}\!\!\left[\hat{R}_{i,m}\!\left(\!{\bm \gamma}^{(m)},\mathcal{H}_{i-1}\!\right)\right]\!\mathbbm{1}\left(i\leq N_T\right)\Big|\mathcal{H}_{i-1}\right]
\end{align}
since $\mathbbm{1}\left(i\leq N_T\right)$ is independent of ${\bm \tau}_i^{(m)}$ given $\mathcal{H}_{i-1}$. We can similarly define the following statistical average of the $i$th epoch length with $m$ update attempts:
\begin{align}
\hat{y}_{i,m}\left({\bm \gamma}^{(m)},\mathcal{H}_{i-1}\right)\triangleq\mathbb{E}\left[y_{i,m}\Big|{\bm \tau}_i^{(m)}={\bm \gamma}^{(m)},\mathcal{H}_{i-1}\right].
\end{align}

Next, observe that by (\ref{eq_aoi_area}) the following holds:
\begin{align} \label{eq_aoi_area_bd}
\hspace{-.1in}\frac{1}{T}\sum_{i=1}^\infty\! R_i\mathbbm{1}\left(i\leq N_T-1\right)\!\leq\!\frac{r(T)}{T}\!\leq\!\frac{1}{T}\sum_{i=1}^\infty\! R_i\mathbbm{1}\left(i\leq N_T\right).
\end{align}
\begin{figure*}
\begin{align}
\frac{1}{T}\mathbb{E}&\left[\sum_{i=1}^\infty R_i\mathbbm{1}\left(i\leq N_T\right)\right] \nonumber \\
\geq&\frac{\mathbb{E}\left[\sum_{i=1}^\infty R_i\mathbbm{1}\left(i\leq N_T\right)\right]}{\mathbb{E}\left[\sum_{i=1}^\infty y_i\mathbbm{1}\left(i\leq N_T\right)\right]} \label{eq_pf_1} \\
=&\frac{\mathbb{E}\left[\sum_{i=1}^\infty \sum_{m=1}^\infty R_{i,m}\prod_{k=1}^{m-1}\mathbbm{1}\left(e_{i,k}\right)\mathbbm{1}\left(e_{i,m}^c\right)\mathbbm{1}\left(i\leq N_T\right)\right]}{\mathbb{E}\left[\sum_{i=1}^\infty y_i\mathbbm{1}\left(i\leq N_T\right)\right]} \label{eq_pf_2} \\
=&\frac{\sum_{i=1}^\infty \sum_{m=1}^\infty q^{m-1}(1-q)\mathbb{E}\left[R_{i,m}\mathbbm{1}\left(i\leq N_T\right)\right]}{\sum_{i=1}^\infty\mathbb{E}\left[y_i\mathbbm{1}\left(i\leq N_T\right)\right]} \label{eq_pf_3} \\
=&\frac{\sum_{i=1}^\infty \sum_{m=1}^\infty q^{m-1}(1-q) \mathbb{E}_{\mathcal{H}_{i-1}}\left[\mathbb{E}_{{\bm \tau}_i^{(m)}}\left[\hat{R}_{i,m}\left({\bm \gamma}^{(m)},\mathcal{H}_{i-1}\right)\right]\mathbbm{1}\left(i\leq N_T\right)\Big|\mathcal{H}_{i-1}\right]}{\sum_{i=1}^\infty\mathbb{E}\left[y_i\mathbbm{1}\left(i\leq N_T\right)\right]} \label{eq_pf_4} \\
=&\frac{\sum_{i=1}^\infty \mathbb{E}_{\mathcal{H}_{i-1}}\left[\sum_{m=1}^\infty q^{m-1}(1-q)\mathbb{E}_{{\bm \tau}_i^{(m)}}\left[\hat{R}_{i,m}\left({\bm \gamma}^{(m)},\mathcal{H}_{i-1}\right)\right]\mathbbm{1}\left(i\leq N_T\right)\Big|\mathcal{H}_{i-1}\right]}{\sum_{i=1}^\infty\mathbb{E}\left[y_i\mathbbm{1}\left(i\leq N_T\right)\right]} \label{eq_pf_5} \\
=&\frac{\sum_{i=1}^\infty \! \mathbb{E}_{\mathcal{H}_{i-1}} \!\! \left[ \sum_{m=1}^\infty \! q^{m-1}(1-q)\mathbb{E}_{{\bm \tau}_i^{(m)}} \!\! \left[\hat{y}_{i,m}\left({\bm \gamma}^{(m)},\mathcal{H}_{i-1}\right)\right] \! \frac{\sum_{m=1}^\infty q^{m-1}(1\!-\!q)\mathbb{E}_{{\bm \tau}_i^{(m)}}\left[\hat{R}_{i,m}\left({\bm \gamma}^{(m)},\mathcal{H}_{i-1}\right)\right]}{\sum_{m=1}^\infty q^{m-1}(1-q)\mathbb{E}_{{\bm \tau}_i^{(m)}}\left[\hat{y}_{i,m}\left({\bm \gamma}^{(m)},\mathcal{H}_{i-1}\right)\right]} \mathbbm{1}\left(i\leq N_T\right) \Big| \mathcal{H}_{i-1}\right]}{\sum_{i=1}^\infty\mathbb{E}\left[y_i\mathbbm{1}\left(i\leq N_T\right)\right]} \label{eq_pf_6} \\
\geq&\frac{\sum_{i=1}^\infty \mathbb{E}_{\mathcal{H}_{i-1}}\left[ \sum_{m=1}^\infty q^{m-1}(1-q)\mathbb{E}_{{\bm \tau}_i^{(m)}}\left[\hat{y}_{i,m}\left({\bm \gamma}^{(m)},\mathcal{H}_{i-1}\right)\right] R^*\left(\mathcal{H}_{i-1}\right) \mathbbm{1}\left(i\leq N_T\right) \Big| \mathcal{H}_{i-1}\right]}{\sum_{i=1}^\infty\mathbb{E}\left[y_i\mathbbm{1}\left(i\leq N_T\right)\right]} \label{eq_pf_7} \\
\geq&\frac{\sum_{i=1}^\infty \sum_{m=1}^\infty q^{m-1}(1-q) \mathbb{E}_{\mathcal{H}_{i-1}}\left[\mathbb{E}_{{\bm \tau}_i^{(m)}}\left[\hat{y}_{i,m}\left({\bm \gamma}^{(m)},\mathcal{H}_{i-1}\right)\right] \mathbbm{1}\left(i\leq N_T\right) \Big| \mathcal{H}_{i-1}\right]}{\sum_{i=1}^\infty\mathbb{E}\left[y_i\mathbbm{1}\left(i\leq N_T\right)\right]} R_{\min} \label{eq_pf_8} \\
=&\frac{\sum_{i=1}^\infty \sum_{m=1}^\infty q^{m-1}(1-q)\mathbb{E}\left[y_{i,m}\mathbbm{1}\left(i\leq N_T\right)\right]}{\sum_{i=1}^\infty\mathbb{E}\left[y_i\mathbbm{1}\left(i\leq N_T\right)\right]} R_{\min} \label{eq_pf_9} \\
=&R_{\min}. \label{eq_pf_10}
\end{align}
\hrulefill
\end{figure*}
\hspace{-.075in}Following similar analysis as in \cite[Appendix C-1]{jing-age-online}, one can show that the term $\mathbb{E}\left[R_{N_T}\right]/T\rightarrow0$ as $T\rightarrow\infty$, making the upper and lower bounds in (\ref{eq_aoi_area_bd}) equal as $T\rightarrow\infty$. Therefore, we proceed by deriving a lower bound on $\frac{1}{T}\mathbb{E}\left[\sum_{i=1}^\infty R_i\mathbbm{1}\left(i\leq N_T\right)\right]$ and conclude that it shall also serve as a lower bound on $\frac{1}{T}\mathbb{E}\left[r(T)\right]$ as $T\rightarrow\infty$ (the objective function of the problem). We do so through a series of inequalities at the top of the next page. There, (\ref{eq_pf_1}) follows since, by definition of $N_T$, it holds that $\mathbb{E}\left[\sum_{i=1}^\infty y_i\mathbbm{1}\left(i\leq N_T\right)\right]\geq T$; (\ref{eq_pf_2}) follows by (\ref{eq_r_rm}); (\ref{eq_pf_3}) follows by the monotone convergence theorem, and the fact that erasure events are mutually independent and are independent of transmissions; (\ref{eq_pf_4}) follows by (\ref{eq_rm_itr_exp}); (\ref{eq_pf_5}) follows again by the monotone convergence theorem; $R^*\left(\mathcal{H}_{i-1}\right)$ in (\ref{eq_pf_7}) denotes the minimum value of $\frac{\sum_{m=1}^\infty q^{m-1}(1-q)\mathbb{E}_{{\bm \tau}_i^{(m)}}\left[\hat{R}_{i,m}\left({\bm \gamma}^{(m)},\mathcal{H}_{i-1}\right)\right]}{\sum_{m=1}^\infty q^{m-1}(1-q)\mathbb{E}_{{\bm \tau}_i^{(m)}}\left[\hat{y}_{i,m}\left({\bm \gamma}^{(m)},\mathcal{H}_{i-1}\right)\right]}$; $R_{\min}$ in (\ref{eq_pf_8}) denotes the minimum value of $R^*\left(\mathcal{H}_{i-1}\right)$ over all epochs and their corresponding histories, i.e., the minimum over all $i$ and $\mathcal{H}_{i-1}$; and (\ref{eq_pf_9}) and (\ref{eq_pf_10}) follow by the relationships between $\hat{y}_{i,m}$, $y_{i,m}$, and $y_i$ which are the same as those between $\hat{R}_{i,m}$, $R_{i,m}$, and $R_i$ that got us from (\ref{eq_pf_1}) to (\ref{eq_pf_4}).

Note that the online policy achieving $R^*\left(\mathcal{H}_{i-1}\right)$ is only a function of the energy arrivals in the $i$th epoch, since the history $\mathcal{H}_{i-1}$ is fixed. Now observe that by the memoryless property of exponential distribution, $\tau_{i,k}$'s are i.i.d.$\sim\exp(1)$. Therefore, if one repeats the policy that achieves $R_{\min}$ over all epochs, which is possible since the genie provides information about epochs' start times, then one gets a renewal policy in which $y_i$'s are i.i.d.

We now argue that the best renewal policy does not depend on the genie's provided information. First, it is clear that when an epoch starts, the sensor's next inter-update attempt becomes independent of the past and only a function of the energy arrivals in the epoch, in particular the first arrival time. If the sensor receives an information from the genie that its first update was successful, then this means a new epoch started and the process is repeated. On the other hand, if it does not hear from the genie, it is not allowed to act upon that information according to our enforced constraint that we stated at the beginning of the proof. Hence, it repeats the same policy, otherwise the constraint would be violated. Therefore, the policy does not change whether the genie sends its information or not. Finally, observe that this policy is achievable in the original system considered in this paper, i.e., the system with no genie. This completes the proof.

\subsection{Proof of Theorem~\ref{thm:threshold_noFB}} \label{apndx_thm:threshold_noFB}

We first evaluate the terms $\mathbb{E}\left[y\right]$ and $\mathbb{E}\left[R\right]$. The expected epoch length can be found using iterated expectations by conditioning on how many erasure events occurred in it. We now write the following:
\begin{align}
\mathbb{E}\left[y\right]=&(1-q)\mathbb{E}\left[x(\tau_1)\right] \nonumber \\
&+q(1-q)\left(\mathbb{E}\left[x(\tau_1)\right]+\mathbb{E}\left[x(\tau_2)\right]\right) \nonumber \\
&+q^2(1-q)\left(\mathbb{E}\left[x(\tau_1)\right]+\mathbb{E}\left[x(\tau_2)\right]+\mathbb{E}\left[x(\tau_3)\right]\right) \nonumber \\
&+\dots \\
=&\mathbb{E}\left[x(\tau)\right]\left(1+q+q^2+\dots\right) \\
=&\frac{\mathbb{E}\left[x(\tau)\right]}{1-q}, \label{eq_no_fb_L}
\end{align}
where $\tau\sim\text{exp}(1)$, and the second equality follows since $\tau_j$'s are i.i.d. $\text{exp}(1)$ random variables by the memoryless property of the exponential distribution. The expected area under the age curve in a single epoch can be found similarly as follows:
\begin{align}
\mathbb{E}\left[R\right]=&(1-q)\frac{1}{2}\mathbb{E}\left[x^2(\tau_1)\right] \nonumber \\
&+q(1-q)\frac{1}{2}\mathbb{E}\left[\left(x(\tau_1)+x(\tau_2)\right)^2\right] \nonumber \\
&+q^2(1-q)\frac{1}{2}\mathbb{E}\left[\left(x(\tau_1)+x(\tau_2)+x(\tau_3)\right)^2\right] \nonumber \\
&+\dots \\
=&\frac{1}{2}\mathbb{E}\left[x^2(\tau)\right]\left(1+q+q^2+\dots\right) \nonumber \\
&+\left(\mathbb{E}\left[x(\tau)\right]\right)^2\left(q+2q^2+3q^3+\dots\right) \\
=&\frac{\frac{1}{2}\mathbb{E}\left[x^2(\tau)\right]}{1-q}+\frac{q\left(\mathbb{E}\left[x(\tau)\right]\right)^2}{(1-q)^2}, \label{eq_no_fb_R}
\end{align}
where the second equality again follows since $\tau_j$'s are i.i.d., and after some algebraic manipulations.

Using (\ref{eq_no_fb_L}) and (\ref{eq_no_fb_R}), one can write the following Lagrangian \cite{boyd} for problem (\ref{opt_no_fb_aux}):
\begin{align}
\mathcal{L}=&\frac{\frac{1}{2}\mathbb{E}\left[x^2(\tau)\right]}{1-q}+\frac{q\left(\mathbb{E}\left[x(\tau)\right]\right)^2}{(1-q)^2}-\lambda\frac{\mathbb{E}\left[x(\tau)\right]}{1-q} 
\nonumber \\
&-\int_0^\infty \left(x(\tau)-\tau\right)\eta(\tau)d\tau,
\end{align}
where $\eta$ is a Lagrange multiplier. Taking (the functional) derivative with respect to $x(t)$ and equating to $0$, we get that the optimal $x$ satisfies
\begin{align}
x(t)=\lambda-\frac{2q}{1-q}\mathbb{E}\left[x(\tau)\right]+\frac{\eta(t)}{e^{-t}/1-q}.
\end{align}
Now let us define
\begin{align} \label{eq_no_fb_lmda_prm_ini}
\lambda^\prime \triangleq \lambda-\frac{2q}{1-q}\mathbb{E}\left[x(\tau)\right].
\end{align}
The sign of $\lambda^\prime$ has a major implication on the optimal policy's structure, which we discuss in detail next.

If $\lambda^\prime<0$ then we must have $\eta(t)>0$, $\forall t$, to maintain positivity of $x(t)$. By complementary slackness \cite{boyd} this further implies that $x(t)=t$, $\forall t$, i.e., a {\it greedy zero-wait policy} is optimal in this case, in which energy is used to send an update whenever it arrives. This case occurs for relatively high values of $q$ which we specify precisely towards the end of this section. The value of $p^{\text{noFB}}(\lambda)$ in this case can be computed by plugging in $x(\tau)=\tau$ with $\mathbb{E}\left[x(\tau)\right]=1$ and $\mathbb{E}\left[x^2(\tau)\right]=2$ to get after some direct manipulations that
\begin{align}
p^{\text{noFB}}(\lambda)=\frac{1-\lambda(1-q)}{(1-q)^2},
\end{align}
which admits an optimal long-term average AoI, $\lambda^*$, of
\begin{align} \label{eq_greedy_aoi}
\lambda^*=\frac{1}{1-q}.
\end{align}
Note that such greedy policy is always feasible and therefore (\ref{eq_greedy_aoi}) can generally serve as an upper bound on $\lambda^*$.

Now if $\lambda^\prime\geq0$, then by complementary slackness \cite{boyd} we get that (see \cite{jing-age-online} and \cite{arafa-age-online-finite})
\begin{align} \label{eq_no_fb_x}
x(t)=\begin{cases}\lambda^\prime,\quad &t<\lambda^\prime\\ t,\quad &t\geq\lambda^\prime \end{cases}.
\end{align}
That is, the optimal status update policy is a $\lambda^\prime$-{\it threshold policy}. Using this, one can directly compute $\mathbb{E}\left[x(\tau)\right]=\lambda^\prime+e^{-\lambda^\prime}$ and substitute back in (\ref{eq_no_fb_lmda_prm_ini}) to get that
\begin{align} \label{eq_no_fb_lmda_prm_apndx}
\frac{1+q}{1-q}\lambda^\prime+\frac{2q}{1-q}e^{-\lambda^\prime}=\lambda.
\end{align}
Direct first derivative analysis shows that the left hand side above is increasing in $\lambda^\prime$ for $\lambda^\prime\geq0$, and therefore, since its value at $\lambda^\prime=0$ is $2q/(1-q)$, (\ref{eq_no_fb_lmda_prm_apndx}) has a unique solution in $\lambda^\prime$ for every given $\lambda\geq2q/(1-q)$, i.e., $2q/(1-q)$ is the best achievable long-term average AoI if $\lambda^\prime\geq0$. Now observe that for $q>1/2$, the greedy zero-wait policy achieves a lower long-term average AoI than that, given by $1/(1-q)$. We therefore conclude that in the optimal policy, $\lambda^\prime$ can only be non-negative if $q\leq1/2$. Continuing with this assumption, we use (\ref{eq_no_fb_x}), and some algebraic manipulations, to get
\begin{align} \label{eq_no_fb_p}
p^{\text{noFB}}(\lambda^\prime)=\frac{(1-q)\left(e^{-\lambda^\prime}-\frac{1}{2}\left(\lambda^\prime\right)^2\right)-q\left(\lambda^\prime+e^{-\lambda^\prime}\right)^2}{(1-q)^2},
\end{align}
with $\lambda^\prime$ as defined in (\ref{eq_no_fb_lmda_prm_apndx}). Now observe that solving $p^{\text{noFB}}\left(\lambda^\prime\right)=0$ for $\lambda^\prime\geq0$ is tantamount to having $p^{\text{noFB}}(0)\geq0$ (since $p^{\text{noFB}}(\lambda)$ is monotonically decreasing \cite{dinkelbach-fractional-prog} in $\lambda$, and $\lambda$ is an increasing function of $\lambda^\prime$ from (\ref{eq_no_fb_lmda_prm_apndx})). In other words, we must have
\begin{align}
p^{\text{noFB}}(0)=\frac{1-2q}{(1-q)^2}\geq0 \iff q\leq\frac{1}{2}
\end{align}
as assumed before.

In conclusion, the optimal policy's structure depends on the value of the erasure probability, $q$. If $q>\frac{1}{2}$ then (\ref{eq_no_fb_p}) does not admit a positive $\lambda^\prime$ solution for $p^{\text{noFB}}\left(\lambda^\prime\right)=0$, and therefore it holds that $\lambda^\prime<0$, and the greedy zero-wait policy is optimal. While if $q\leq\frac{1}{2}$ then the optimal policy is a $\lambda^\prime$-threshold policy as in (\ref{eq_no_fb_x}), with the optimal $\lambda^\prime$ solving $p^{\text{noFB}}\left(\lambda^\prime\right)=0$.

\subsection{Proof of Lemma~\ref{thm:thrshld_grd_fb}} \label{apndx_thm:thrshld_grd_fb}

First, we prove the direct part: If $x_i$, $i\geq2$, are all greedy policies, then the optimal $x_1$ is a $\gamma$-threshold policy with $\gamma=\left[\lambda-\frac{q}{1-q}\right]^+$. We start by the simplifying the expected epoch length as follows:
\begin{align}
\mathbb{E}\left[y\left({\bm x}\right)\right]=&(1-q)\mathbb{E}\left[x_1(\tau_1)\right]+q(1-q)\left(1+\mathbb{E}\left[x_1(\tau_1)\right]\right) \nonumber \\
&+q^2(1-q)\left(2+\mathbb{E}\left[x_1(\tau_1)\right]\right)+\dots \nonumber \\
&+q^{i-1}(1-q)\left(i-1+\mathbb{E}\left[x_1(\tau_1)\right]\right)+\dots \\
=&\mathbb{E}\left[x_1(\tau_1)\right]+\frac{q}{1-q}. \label{eq_L_final_fb}
\end{align}
Before simplifying the expected epoch reward, let us define $G_i\triangleq\sum_{j=2}^i\tau_j$, $i\geq2$. We now proceed as follows:
\begin{align}
&\hspace{-.15in}\mathbb{E}\left[R\left({\bm x}\right)\right] \nonumber \\
=&(1-q)\frac{1}{2}\mathbb{E}\left[x_1^2(\tau_1)\right] +\sum_{i=2}^\infty q^{i-1}(1-q)\mathbb{E}\left[\left(G_i+x_1(\tau_1)\right)^2\right] \\
=&(1-q)\frac{1}{2}\mathbb{E}\left[x_1^2(\tau_1)\right] \nonumber \\
&+\sum_{i=2}^\infty q^{i-1}(1-q)\mathbb{E}\left(\frac{1}{2}\mathbb{E}\left[G_i^2\right]+\frac{1}{2}\mathbb{E}\left[x_1^2(\tau_1)\right]\right. \nonumber \\
&\hspace{1.75in}+\mathbb{E}\left[G_i\right]\mathbb{E}\left[x_1(\tau_1)\right]\bigg) \\
=&\frac{1}{2}\mathbb{E}\left[x_1^2(\tau_1)\right]+\frac{q}{1-q}\mathbb{E}\left[x_1(\tau_1)\right] \nonumber \\
&+\frac{1}{2}\sum_{i=1}^\infty\left(i-1+(i-1)^2\right)q^{i-1}(1-q) \label{eq_smplfy_R_fb} \\
=&\frac{1}{2}\mathbb{E}\left[x_1^2(\tau_1)\right]+\frac{q}{1-q}\mathbb{E}\left[x_1(\tau_1)\right]+\frac{q}{(1-q)^2}, \label{eq_R_final_fb}
\end{align}
where (\ref{eq_smplfy_R_fb}) follows by the fact that that $G_i$ has a gamma distribution with parameters $i-1$ and $1$, and, in particular, its second moment is given by $\mathbb{E}\left[G_i^2\right]=i-1+(i-1)^2$.

We now plug (\ref{eq_L_final_fb}) and (\ref{eq_R_final_fb}) into the objective function of problem (\ref{opt_aux_fb}), and introduce the following Lagrangian \cite{boyd}:
\begin{align}
\mathcal{L}=&\frac{1}{2}\mathbb{E}\left[x_1^2(\tau_1)\right]+\left(\frac{q}{1-q}-\lambda\right)\mathbb{E}\left[x_1(\tau_1)\right]+\frac{q}{(1-q)^2} \nonumber \\
&-\lambda\frac{q}{1-q}-\int_0^\infty\eta_1(\tau_1)\left(x_1(\tau_1)-\tau_1\right)d\tau_1,
\end{align}
where $\eta_1$ is a Lagrange multiplier. Taking the (functional) derivative with respect to $x_1(t)$ and equating to $0$ we get
\begin{align}
x_1(t)=\left(\lambda-\frac{q}{1-q}\right)+\frac{\eta_1(t)}{e^{-t}}.
\end{align}
We now have two cases. The first is when $\lambda<\frac{q}{1-q}$, whence $\eta_1(t)$ must be strictly positive $\forall t$, which implies by complementary slackness \cite{boyd} that $x_1(t)=t,~\forall t$. In other words, $x_1$ in this case is a greedy policy, or equivalently a $0$-threshold policy. The second case is when $\lambda\geq\frac{q}{1-q}$, in which similar analysis to that in the proof of Theorem~\ref{thm:threshold_noFB} (see also \cite[Section~3]{arafa-age-online-finite}) can be carried out to show that $x_1$ is a $\left(\lambda-\frac{q}{1-q}\right)$-threshold policy. Combining both cases concludes the proof of the direct part.

We now prove the converse part: if the optimal $x_1$ is a $\gamma$-threshold policy, then the optimal $x_i$, $i\geq2$, are all greedy policies. Hence, the first update attempt occurs optimally (by hypothesis) at $x_1(\tau_1)$. Assume that it fails. Note that, by construction, $\tau_2>x_1(\tau_1)$ (see Fig.~\ref{fig:tau_epoch_fb}). Let $s_2\triangleq\tau_2+x_1(\tau_1)$, and let $x_2$ be {\it not} greedy: $x_2(s_2)=s_2^\prime$ for some $s_2^\prime>s_2$. Now consider a slightly different energy arrival pattern, in which the first energy arrival occurs at $s_2$, as opposed to $\tau_1$. Since $s_2>x_1(\tau_1)$, and $x_1$ is an optimal threshold policy, therefore it holds that $x_1(s_2)=s_2$, i.e., it is optimal to update right away at time $s_2$ in the second sample path situation.

Now observe that in both situations the AoI $a(s_2)=s_2$; and, by the memoryless property of exponential distribution, that the time until the next energy arrival after $s_2$ is $\sim\exp(1)$. In addition, the probability that an update gets erased is independent of past erasures. Given that $a(s_2)=s_2$, the upcoming energy arrival is $\sim\exp(1)$, and the probability of erasure is $q$, the optimal decision in the second situation is $x_1(s_2)=s_2$, i.e., update exactly at $s_2$. Therefore, in the first situation, in which the same statistical conditions hold at $s_2$, it {\it cannot be optimal} to wait and update at time $s_2^\prime$. Hence, $x_2$ must be greedy. Similar arguments hold to show that $x_i$, $i\geq3$, must all be greedy as well, given that the optimal $x_1$ is a threshold policy. This concludes the proof of the converse part, and that of the lemma.

\subsection{Proof of Theorem~\ref{thm:threshold_wFB}} \label{apndx_thm:threshold_wFB}

We start by substituting $x_1$ into equations (\ref{eq_L_final_fb}) and (\ref{eq_R_final_fb}) (see Appendix~\ref{apndx_thm:thrshld_grd_fb}) for two cases. First, for $\lambda<\frac{q}{1-q}$, $x_1$ is greedy, i.e., $\mathbb{E}\left[x_1(\tau_1)\right]=1$ and $\mathbb{E}\left[x_1^2(\tau_1)\right]=2$. Therefore, $p^{\text{wFB}}(\lambda)=1-\lambda\frac{1}{1-q}+\frac{2q-q^2}{(1-q)^2}$. Second, for $\lambda\geq\frac{q}{1-q}$, $x_1$ is a $\left(\lambda-\frac{q}{1-q}\right)$-threshold policy, and by direct computation $\mathbb{E}\left[x_1(\tau_1)\right]=\frac{1}{2}\left(\lambda-\frac{q}{1-q}\right)$ and $\mathbb{E}\left[x_1^2(\tau_1)\right]=2e^{-\left(\lambda-\frac{q}{1-q}\right)}$. Therefore, $p^{\text{wFB}}(\lambda)=e^{-\left(\lambda-\frac{q}{1-q}\right)}-\frac{1}{2}\lambda^2+\frac{2q-q^2}{2(1-q)^2}$. In summary, we have
\begin{align}
p^{\text{wFB}}(\lambda)=\begin{cases}1-\lambda\frac{1}{1-q}+\frac{2q-q^2}{(1-q)^2},\quad&\lambda<\frac{q}{1-q}\\
e^{-\left(\lambda-\frac{q}{1-q}\right)}-\frac{1}{2}\lambda^2+\frac{2q-q^2}{2(1-q)^2},\quad&\lambda\geq\frac{q}{1-q}\end{cases}.
\end{align}

We now find $\lambda^*$ that solves $p^{\text{wFB}}(\lambda^*)=0$. It can be directly checked that for $\lambda<\frac{q}{1-q}$, $p^{\text{wFB}}(\lambda)=1-\lambda\frac{1}{1-q}+\frac{2q-q^2}{(1-q)^2}>0$. Thus, focusing on the case $\lambda\geq\frac{q}{1-q}$, $\lambda^*$ is found by solving
\begin{align} \label{eq_lmda_fb_apndx}
e^{-\left(\lambda^*-\frac{q}{1-q}\right)}+\frac{2q-q^2}{2(1-q)^2}=\frac{1}{2}\left(\lambda^*\right)^2,
\end{align}
which admits a unique solution that is strictly larger than $\frac{q}{1-q}$. This can be readily verified by observing that, $1)$ the right hand side of (\ref{eq_lmda_fb_apndx}) is smaller than the left hand side for $\lambda^*=q/(1-q)$; and $2)$ the right hand side of (\ref{eq_lmda_fb_apndx}) is increasing in $\lambda^*$ while the left hand side is decreasing.

\subsection{Proof of Theorem~\ref{thm:multisrc_noFB}} \label{apndx_thm:multisrc_noFB}

Let us define an epoch with respect to some source $j$ as the time elapsed in between two consecutive successful updates of which. Since we focus on RR scheduling, we let $\tau(j)$ denote the time until the energy unit dedicated to source $j$ arrives {\it following the transmission attempt of source $j-1$.} A $\gamma$-threshold policy therefore indicates that source $j$ is sampled after $\max\{\gamma,\tau(j)\}$ time units following the transmission attempt of source $j-1$. Such policy is repeated since the sensor does not know the erasure status, and is identical across all sources. Therefore, it takes
\begin{align}
\sum_{j=1}^M\max\{\gamma,\tau(j)\}
\end{align}
time units to finish {\it one round} of update attempts for all the sources.

Note that a successful transmission for source $j$ may require multiple rounds, and so we denote the total epoch length for source $j$ by $\alpha(j)$. Hence, the long-term average AoI for source $j$ is given by
\begin{align} \label{eq_aoi_muti_nofb}
\frac{\frac{1}{2}\mathbb{E}\left[\alpha(j)^2\right]}{\mathbb{E}\left[\alpha(j)\right]},
\end{align}
which is also identical across sources. Therefore, $\tilde{\rho}_{q,M}^{\text{noFB}}\left(\text{RR},\gamma\right)$ is given by the expression in (\ref{eq_aoi_muti_nofb}), which we evaluate next to prove the theorem.

Since $\tau(j)$'s are i.i.d.$\sim\exp(1)$, we deduce that
\begin{align}
\mathbb{E}\left[\max\{\tau(j),\gamma\}\right]=&\gamma+e^{-\gamma},\quad\forall j, \\
\mathbb{E}\left[\max\{\tau(j),\gamma\}^2\right]=&\gamma^2+2(\gamma+1)e^{-\gamma}, \quad\forall j.
\end{align}
One can now follow similar analysis to that in the derivations of (\ref{eq_no_fb_L}) and (\ref{eq_no_fb_R}) to conclude that
\begin{align}
\mathbb{E}\left[\alpha(j)\right]=&\frac{M\left(\gamma+e^{-\gamma}\right)}{1-q}, \label{eq_aoi_mutisrc_nofb_1} \\
\mathbb{E}\left[\alpha(j)^2\right]=&\frac{M\left(\gamma^2+2(\gamma+1)e^{-\gamma}\right)+M(M\!-\!1)\left(\gamma+e^{-\gamma}\right)^2}{1-q} \nonumber \\
&+\frac{2M^2\left(\gamma+e^{-\gamma}\right)^2q}{(1-q)^2}. \label{eq_aoi_mutisrc_nofb_2}
\end{align}

Substituting (\ref{eq_aoi_mutisrc_nofb_1}) and (\ref{eq_aoi_mutisrc_nofb_2}) into (\ref{eq_aoi_muti_nofb}) gives (\ref{eq_rho_M_noFB}) and concludes the proof.

\subsection{Proof of Theorem~\ref{thm:multisrc_wFB}} \label{apndx_thm:multisrc_wFB}

We follow a similar approach as that in Appendix~\ref{apndx_thm:multisrc_noFB} with a slight change of notation due to the presence of feedback. Specifically, we now define $\alpha(j)$ as the time needed for source $j$ to finish its successful transmission {\it starting from the point at which its turn comes up,} and $\alpha\left(\overline{j}\right)$ as the time needed for the other sources $[M]\backslash j$ to finish their successful transmissions. Since we focus on MAF scheduling, an epoch with respect to source $j$ takes
\begin{align}
\alpha\left(\overline{j}\right)+\alpha(j)   
\end{align}
time units to finish. Hence, the long-term average AoI for source $j$ is given by
\begin{align} \label{eq_aoi_muti_wfb}
\frac{\frac{1}{2}\mathbb{E}\left[\left(\alpha\left(\overline{j}\right)+\alpha(j)\right)^2\right]}{\mathbb{E}\left[\alpha\left(\overline{j}\right)+\alpha(j)\right]},
\end{align}
which is also identical across sources. Therefore, $\tilde{\rho}_{q,M}^{\text{wFB}}\left(\text{MAF},\gamma\right)$ is given by the expression in (\ref{eq_aoi_muti_wfb}), which we evaluate next to prove the theorem.

Let $\tau_1(j)$ denote the time until the first energy unit in the epoch allocated to source $j$ arrives. A $\gamma$-threshold policy therefore implies that the sensor samples source $j$ for the first time after $\max\{\gamma,\tau_1(j)\}$ time units from the time its turn comes up. If that first update is not successful, the sensor continues in a greedy manner; it waits for another energy arrival, which takes $\tau_2(j)$ time units to arrive, and then immediately re-samples source $j$ for a re-transmission. This continues until source $j$'s update is received. If source $j$'s update is received at the $i$th attempt, then
\begin{align}
\alpha(j)=\max\{\gamma,\tau_1(j)\}+\sum_{k=1}^i\tau_k(j),
\end{align}
which occurs with probability $q^{i-1}(1-q)$. Since $\tau_k(j)$'s are i.i.d. $\sim\exp(1)$, one can show that
\begin{align} \label{eq_thm_mltisrc_1}
\mathbb{E}\left[\alpha(j)\right]=\gamma+e^{-\gamma}+\frac{q}{1-q}
\end{align}
in a manner similar to showing (\ref{eq_L_final_fb}). Next, we note that
\begin{align}
\alpha\left(\overline{j}\right)=\sum_{l\in[M]\backslash j}\alpha(l).
\end{align}
Therefore, one can use the symmetry across the sources and write
\begin{align} \label{eq_thm_mltisrc_2}
\mathbb{E}\left[\alpha\left(\overline{j}\right)\right]=(M-1)\left(\gamma+e^{-\gamma}+\frac{q}{1-q}\right).
\end{align}

Next, one can follow similar arguments as above, and as carried out in the analysis of (\ref{eq_R_final_fb}) to evaluate the second moment of $\alpha(j)$ as
\begin{align} \label{eq_thm_mltisrc_3}
\mathbb{E}\left[\alpha(j)^2\right]=&\left(\gamma^2+2(\gamma+1)e^{-\gamma}\right)+2\left(\gamma+e^{-\gamma}\right)\frac{q}{1-q} \nonumber \\
&+\frac{2q}{(1-q)^2}.
\end{align}
Finally, one can invoke the symmetry among the sources again to show that
\begin{align} \label{eq_thm_mltisrc_4}
\mathbb{E}\left[\alpha\left(\overline{j}\right)^2\right]=&(M-1)(M-2)\left(\gamma+e^{-\gamma}+\frac{q}{1-q}\right)^2\nonumber \\
&+(M-1)\left(\gamma^2+2(\gamma+1)e^{-\gamma}\right) \nonumber \\
&+2\left(\gamma+e^{-\gamma}\right)\frac{q}{1-q}+\frac{2q}{(1-q)^2}.
\end{align}

The proof is concluded by plugging (\ref{eq_thm_mltisrc_1}), (\ref{eq_thm_mltisrc_2}), (\ref{eq_thm_mltisrc_3}) and (\ref{eq_thm_mltisrc_4}) into (\ref{eq_aoi_muti_wfb}), using the fact that $\alpha(j)$ and $\alpha\left(\overline{j}\right)$ are independent, and simple rearrangements.


\end{document}